\documentclass[reqno]{amsart} 
\pdfoutput=1

\usepackage{graphicx}

\def \V{\mathbb{V}}
\def\II{\hbox{{1}\kern-.25em\hbox{l}}}

\numberwithin{equation}{section}
\def\stackreb#1#2{\ \mathrel{\mathop{#1}\limits_{#2}}}
\newcommand{\nc}{\newcommand}
\newcommand{\Z}{\mathbb Z}
\nc{\rnc}{\renewcommand} \nc{\beq}{\begin{equation}}
\nc{\eeq}{\end{equation}} \nc{\beqa}{\begin{eqnarray}}
\nc{\eeqa}{\end{eqnarray}}

\def\stackreb#1#2{\ \mathrel{\mathop{#1}\limits_{#2}}}

\begin{document}
\rightline{\vbox{\small\hbox{\tt HU-EP-15-24} }}
 \vskip 2.2 cm

\title[Star-triangle relation and $3d$ indices]
{\bf The star-triangle relation and \\ $3d$ superconformal indices}

\author{I. Gahramanov}\address{Institut f\"{u}r Physik und IRIS Adlershof, Humboldt-Universit\"{a}t zu Berlin,
Zum Grossen Windkanal 6, D12489 Berlin, Germany and Institute of Radiation Problems ANAS, B.Vahabzade 9, AZ1143 Baku, Azerbaijan}
\email{ilmar@physik.hu-berlin.de}

\author{V. P. Spiridonov}\address{Bogoliubov Laboratory of Theoretical Physics, JINR,
 Dubna, Moscow reg. 141980, Russia}
 \email{spiridon@theor.jinr.ru }

\begin{abstract}
Superconformal indices  of $3d$ $\mathcal{N }=2$  supersymmetric
field theories  are investigated from the Yang-Baxter equation point of view.
Solutions of the star-triangle relation, vertex and IRF Yang-Baxter equations
are expressed in terms of the $q$-special functions associated with
these $3d$ indices. For a two-dimensional monopole-spin system
on the square lattice a free energy per spin is explicitly determined.
Similar to the partition functions, superconformal indices of $3d$ theories with
the chiral symmetry breaking reduce to Dirac delta functions with the support
on chemical potentials of the preserved flavor groups.
\end{abstract}

\maketitle

\tableofcontents

\section{Introduction}

Special functions \cite{aar} are key mathematical objects in solvable models
of physical phenomena. Quantum integrable systems and related
Yang-Baxter equations and quantum algebras \cite{Baxter,Faddeev,Jimbo,TF}
have been investigated for a long time in relation to plain hypergeometric
functions, their $q$-analogues and elliptic functions. Fairly recently
the third class of transcendental functions of hypergeometric type
called elliptic hypergeometric integrals has been discovered \cite{S3},
which strongly extended the database of classical special functions.
The cornerstone of the latter functions is the following elliptic
beta integral

\textbf{Theorem} (Spiridonov \cite{S1}). Let $t_1, \dots ,t_6,p,q \in {\mathbb{C}}$ with $|t_1|, \dots , |t_6|,|p|,|q| <1$ and $\prod_{j=1}^6 t_j=pq$. Then
\begin{equation} \label{Betaint}
\frac{(p;p)_\infty (q;q)_\infty}{2} \int_{\mathbb{T}} \frac{\prod_{i=1}^6
\Gamma(t_i z^{\pm 1} ;p,q)}{\Gamma(z^{\pm 2};p,q)} \frac{dz}{2 \pi \textup{i} z}
= \prod_{1 \leq i < j \leq 6} \Gamma(t_i t_j;p,q),
\end{equation}
where $\Gamma(z ;p,q)=(pqz^{-1};p,q)_\infty/(z;p,q)_\infty$, $(z;p,q)_\infty
=\prod_{j,k=0}^\infty
(1-zp^jq^k)$, is the elliptic gamma function and $\mathbb T$ is the unit circle of
positive orientation.

The first physical application of elliptic hypergeometric integrals
consisted in the interpretation of some of them as wave functions or
normalizations of wave functions in particular quantum mechanical problems  \cite{S3}.
The most important known application of identity \eqref{Betaint} was found in
\cite{Dolan:2008qi} in the context of ${\mathcal N}=1$  supersymmetric field
theories within which it has the meaning of the equality of superconformal indices
 \cite{Kinney,Romelsberger1,Romelsberger2} in Seiberg dual theories
\cite{Seiberg,Seiberg:d}. Indeed, the integral on the left-hand side of
the equality (\ref{Betaint}) is the superconformal index
of the $4d$ ${\mathcal N}=1$ supersymmetric
gauge theory with $SU(2)$ gauge group and $N_F=6$ flavors, chiral
scalar multiplets in the fundamental representation of the flavor
group  $SU(6)$, while the expression on the right side is the superconformal
index for the dual theory without gauge degrees of freedom and the
chiral fields in the 15-dimensional totally antisymmetric tensor representation
of the same flavor group. In other words, the elliptic beta integral is the
manifestation of the $s$-confinement phenomenon in gauge theories \cite{Seiberg}.
The superconformal indices techniques is the most convenient tool
for searching new Seiberg dualities \cite{SV1,SV3,SV3b}. Using properties of
the elliptic hypergeometric integrals one can describe uniformly
the 't Hooft anomaly matching conditions \cite{SV5} and the chiral symmetry
breaking \cite{SV6}. A direct consequence of formula \eqref{Betaint}
was used in topological field theories as well \cite{RR}.

Another application of relation \eqref{Betaint} has lead to important progress
in the study of exactly solvable models of statistical mechanics.
Namely, it has been shown to yield new solutions of the
star-triangle relations either in functional  \cite{BS} or operator forms
\cite{DS}. Actually, the latter form of the star-triangle relation has been found
long before as the integral Bailey lemma \cite{spi:bailey}.
Using the results of \cite{BS}, a correspondence between the quiver gauge
theories and integrable lattice models such that the integrability emerges as
a manifestation of the Seiberg type dualities has been established in \cite{astr}.

Degenerations of the $2d$ spin system of \cite{BS} lead to many known models. For instance,
the Faddeev-Volkov model \cite{VF,BMS} or its
extension \cite{astr} can be obtained in this context as follows.
One can reduce superconformal indices of $4d$ theories to the partition functions
of $3d$ ${\mathcal N}=2$ theories \cite{DSV}. This reduction leads to the equality
of partition functions on the squashed sphere \cite{Hama2} of dual theories
expressed in terms of the hyperbolic hypergeometric integral identities.

The star-triangle relation represents a particular form of the
Yang-Baxter equations (YBE) standing behind the quantum integrable systems.
Another form is the vertex type YBE associated with the integrable spin chains.
A powerful techniques for solving such type of equations was developed in
\cite{SD,DM}. The elliptic beta integral \eqref{Betaint} and related Bailey lemma
\cite{spi:bailey} played a prominent role in building the most complicated
known integral operator solutions of the YBE  \cite{DS}. In particular, this
approach has lead to a new rich class of finite-dimensional
solutions of the YBE \cite{CDS}.

In this paper, we present a new solution of the star-triangle relation
and other forms of YBE
in terms of the basic hypergeometric identity presented in \cite{Ros1}.
We relate the Yang-Baxter equations to three-dimensional supersymmetric
dualities. The new solution corresponds to the generalized superconformal index
of certain $3d$ ${\mathcal N}=2$ superconformal gauge theory
having a distinguished form due to the contribution of monopoles
\cite{Imamura:2011su,Kapustin:2011jm,Kim:2009wb,Krattenthaler:2011da}.
Detailed presentation of this correspondence is given in the last
section.

\section{Notation and definitions}

For $q, z\in\mathbb{C}$, $|q|<1,$ we define the infinite $q$-product
\begin{equation}
(z;q)_\infty \ := \ \prod_{k=0}^\infty (1-z q^k).
\end{equation}
The (normalized) $q$-gamma function of Jackson has the form \cite{aar}
\begin{equation}
\Gamma(z;q) \ :=\frac{1}{(z;q)_\infty}.
\end{equation}
Denote
\begin{equation}
(a,b;q)_\infty:=(a;q)_\infty (b;q)_\infty, \quad
(ax^{\pm1};q)_\infty:=(ax;q)_\infty(ax^{-1};q)_\infty
\end{equation}
with a similar convention for other generalized gamma functions in
\eqref{Betaint} and other relations below.

We need the following $q$-hypergeometric identity.

{\bf Theorem.} (Rosengren  \cite{Ros1}). Let $a_1,\ldots, a_6, q\in\mathbb{C}$
and integers $N_1,\ldots, N_6\in\mathbb Z$, satisfy the constraints $|a_j|, |q|<1$, and
$\prod_{j=1}^6 a_j=q$, $\sum_{j=1}^6N_j=0$. Then
\begin{align} \nonumber
& \sum_{m\in\mathbb{Z}}\int_{\mathbb T}\prod_{j=1}^6
\frac{(q^{1+\frac{m}{2} }\frac{1}{a_jz},q^{1-\frac{m}{2} }\frac{z}{a_j};q)_\infty}
{(q^{N_j+\frac{m}{2} }a_jz,q^{N_j-\frac{m}{2} }\frac{a_j}{z};q)_\infty}
\frac{(1-q^mz^2)(1-q^mz^{-2})}{q^mz^{6m}}\frac{dz}{2\pi iz} \\ \label{qbetaint}
& \qquad =\frac{2}{\prod_{j=1}^6q^{\binom{N_j}{2}}a_j^{N_j}}
\prod_{1\leq j<k\leq 6} \frac{(qa_j^{-1}a_k^{-1};q)_\infty}
{(q^{N_j+N_k}a_ja_k;q)_\infty},
\end{align}
where $\mathbb T$ is the unit circle of positive orientation.

This is a $q$-beta sum-integral associated with $3d$ superconformal indices.
The proof of the theorem is presented in \cite{Ros2}.

Let us define the following generalized $q$-gamma function as a
combination of four $q$-gamma functions and $z^m$ and $a^m$:
\begin{equation}
\Gamma_q(a,n;z,m):=\frac{(q^{1+\frac{n+m}{2} }\frac{1}{az},q^{1+\frac{n-m}{2} }
\frac{z}{a};q)_\infty} {a^nz^m(q^{\frac{n+m}{2}}az,q^{\frac{n-m}{2}}\frac{a}{z};q)_\infty},
\end{equation}
where $a, z\in\mathbb{C}$ and $n, m\in\mathbb{Z}$.

{\bf Lemma.} One has the following inversion relation:
\begin{equation}
\Gamma_q(a,n;z,m)\Gamma_q(b,-n;z,m)=1, \quad ab=q.
\end{equation}
{\bf Proof.}
Consider the explicit form of the indicated product of $\Gamma_q$-functions
after the substitution $b=q/a$:
\begin{eqnarray} \nonumber &&
\Gamma_q(a,n;z,m)\Gamma_q({\textstyle\frac{q}{a}},-n;z,m)
\\ && \makebox[4em]{}
=\frac{q^n}{z^{2m}a^{2n}}
\frac{(q^{1+\frac{n+m}{2} }\frac{1}{az},q^{1+\frac{n-m}{2} }\frac{z}{a},
q^{\frac{-n+m}{2}}\frac{a}{z},q^{\frac{-n-m}{2}}az;q)_\infty}
{(q^{\frac{n+m}{2}}az,q^{\frac{n-m}{2}}\frac{a}{z},
q^{1+\frac{-n+m}{2}}\frac{z}{a},q^{1+\frac{-n-m}{2}}\frac{1}{az};q)_\infty}.
\end{eqnarray}
Using the relation $(a;q)_\infty=(1-a)(aq;q)_\infty$, for $n>m>0$ we can rewrite this
expression as
\begin{equation}
\frac{q^n}{z^{2m}a^{2n}}
\prod_{i=0}^{n+m-1}\frac{ 1-az q^{i-(m+n)/2} }{ 1-a^{-1}z^{-1}q^{i+1-(m+n)/2} }
\prod_{j=0}^{n-m-1}\frac{ 1-a^{-1}zq^{i+1+(n-m)/2} }
{1-az^{-1}q^{i+(n-m)/2} }=1.
\end{equation}
For other possible values of the integers $n$ and $m$ one gets the same
result due to the properties of $q$-Pochhammer symbols.

Now we can rewrite the above $q$-beta sum-integral in the following compact form.

\begin{equation}
\sum_{m\in\mathbb{Z}}\int_{\mathbb T}\prod_{j=1}^6\Gamma_q(a_j,n_j;z,m)[d_mz]
=\frac{1}{\prod_{j=1}^6a_j^{2n_j}}
\prod_{1\leq j<k\leq 6} \frac{(q^{1+\frac{n_j+n_k}{2} }a_j^{-1}a_k^{-1};q)_\infty}
{(q^{ \frac{n_j+n_k}{2} }a_ja_k;q)_\infty},
\label{qbeta}\end{equation}
where $\prod_{j=1}^6 a_j=q$, $\sum_{j=1}^6n_j=0$, and
$$
[d_mz] :=\frac{(1-q^mz^2)(1-q^mz^{-2})}{q^m}\frac{dz}{4\pi iz},\qquad
[d_mz]=[d_{-m}z].
$$

\section{Bailey lemma and the star-triangle relation}

Let us define the $D$-function
\begin{equation}
D(t;a,n;z,m) :=  \Gamma_q (q^{\frac{1}{2}} t^{-1}a, n;z,m)
\Gamma_q (q^{\frac{1}{2}} t^{-1}a^{-1},-n;z,m).
\end{equation}
It is easy to see that
\begin{equation}
D(t^{-1};a,n;z,m)=\frac{1}{D(t;a,n;z,m)}, \qquad  D(1;a,n;z,m)=1.
\end{equation}

Introduce the integral-sum operator of the form
\begin{equation}\label{Mop}
M(t)_{x,n;z,m} f_m(z):=   \frac{(t^2;q)}{(qt^{-2};q)}\;\sum_{m\in\mathbb{Z}}
\int_{\mathbb T} [d_mz] \; \Gamma_{q}(tx^{\pm1},\pm n;z, m) f_m(z),
\end{equation}
where
\begin{align} \nonumber
\Gamma_{q}(tx^{\pm1},\pm n;z, m) : & =\Gamma_{q}(tx,n;z, m) \Gamma_{q}(tx^{-1},-n;z,m) \\
& =D(q^{1/2}t^{-1};x,n;z,m)
\end{align}
and $f_m(z)$ is an arbitrary sequence of holomorphic functions.

We note that the following permutational symmetries hold true
\begin{equation}
\Gamma_{q}(tx^{\pm1},\pm n;z, m)=\Gamma_q(tz^{\pm1},\pm m;x,n),
\label{permsym}\end{equation}
\begin{equation}
D(t;a,n;z,m) = D(t;z,m;a,n).
\end{equation}

Following the original integral generalization \cite{spi:bailey,S3} of the Bailey
chains techniques \cite{aar}, we introduce the notion of Bailey pairs in the
present context.

{\bf Definition.} We say that two sequences of functions $\alpha_m(z;t)$
and $\beta_m(z;t)$, $m\in\mathbb{Z}$, of complex variables $z$ and $t$ form a Bailey pair
with respect to the parameter $t$ if they are related by the integral-sum
transform \eqref{Mop},
\begin{equation}
\beta_n(x;t)=M(t)_{x,n;z,m} \alpha_m(z;t).
\label{BP}\end{equation}
Here we assume that $|tx|,|t/x|<1$ and other regions of parameters are
reached by the analytical continuation.

{\bf Bailey lemma.} Suppose we have a particular Bailey pair $\alpha_k(x;t), \beta_k(x;t)$
with respect to the parameter $t$. Then the sequences of functions
\begin{eqnarray} \label{aBP} &&
\alpha'_k(x;st)= D(s;y,l;x,k)\alpha_k(x;t),
\\ &&
\beta'_k(x;st)=  D(t^{-1};y,l;x,k)M(s)_{x,k;z,m}D(st;y,l;z,m) \beta_m(z;t),
\label{bBP}\end{eqnarray}
where $s, y\in \mathbb{C}, l\in\mathbb{Z}$ are arbitrary new parameters,
form a Bailey pair with respect to the parameter $st$.

{\bf Proof.} Let us substitute primed sequences into the relation
\begin{equation}
\beta_k'(w;st)=M(st)_{w,k;x,j} \alpha_j'(x;st)
\end{equation}
and use the inversion
$D(t^{-1};y,l;x,k)=1/D(t;y,l;x,k)$. This yields the operator identity
\begin{equation}
M(s)_{w,k;z,m} \; D(st;y,l;z,m) M(t)_{z,m;x,j} = D(t;y,l;w,k)M(st)_{w,k;x,j} D(s;y,l;x,j)
\label{STR}\end{equation}
known as the star-triangle relation. It is a straightforward consequence
of the Rosengren $q$-beta sum-integral. First we compute the expression
on the left-hand side of \eqref{STR}
\begin{eqnarray} \nonumber && \makebox[-2em]{}
  \frac{(s^2,t^2;q)}{(qs^{-2},qt^{-2};q)}\sum_{m\in\mathbb Z}\int_{\mathbb T} [d_mz] \;
\Gamma_{q}(sw^{\pm1},\pm k;z,m)\Gamma_{q}(q^{\frac{1}{2}}(st)^{-1}y^{\pm1},\pm l;z,m) \\  \nonumber && \makebox[-2em]{} \times
\sum_{j\in\mathbb Z}  \int_{\mathbb T} [d_jx]
\times  \Gamma_{q}(tz^{\pm1},\pm m; x,j) \\  && \makebox[-2em]{}
=  \frac{(s^2,t^2;q)}{(qs^{-2},qt^{-2};q)} \sum_{j\in\mathbb Z} \int_{\mathbb T} [d_jx]
 \sum_{m\in\mathbb Z} \int_{\mathbb T}\prod_{j=1}^6\Gamma_q(a_j,n_j;z,m)[d_mz],
\end{eqnarray}
where we used the permutational symmetry of $\Gamma_q$-function and have denoted
\begin{eqnarray}\nonumber &&
a_1=sw, \quad n_1=k, \quad  a_2= \frac{s}{w},\quad  n_2=-k, \quad
a_3=\frac{q^{1/2}y}{st}, \quad n_3=l, \;
\\  &&
a_4=\frac{q^{1/2}}{sty},\quad  n_4=-l, \quad  a_5=tx, \quad n_5=j, \quad
a_6=\frac{t}{x},\quad  n_6=-j.
 \end{eqnarray}
The balancing condition holds true $\prod_{j=1}^6a_j=q$, $\sum_{j=1}^6n_j=0$,
and we can apply
the above formula \eqref{qbeta} for computing the
integral over measure $[d_mz]$.
This yields the expression
\begin{eqnarray} \nonumber && \makebox[2em]{}
\frac{ (q^{\frac{1+k+l}{2}}\frac{t}{wy}, q^{\frac{1+k-l}{2}}\frac{ty}{w},
q^{\frac{1-k+l}{2}}\frac{tw}{y},q^{\frac{1-k-l}{2}}twy;q)}
{w^{2k}y^{2l}(q^{\frac{1+k+l}{2}}\frac{wy}{t}, q^{\frac{1+k-l}{2}}\frac{w}{ty},
q^{\frac{1-k+l}{2}}\frac{y}{tw},q^{\frac{1-k-l}{2}}\frac{1}{twy};q) }
\\ \nonumber && \makebox[-2em]{} \times
  \frac{(s^2t^2;q)}{(qs^{-2}t^{-2};q)} \sum_{j\in\mathbb Z} \int_{\mathbb T}
[d_jx]
\frac{ (q^{1+\frac{k+j}{2}}\frac{1}{stwx}, q^{1+\frac{k-j}{2}}\frac{x}{stw},
q^{1+\frac{-k+j}{2}}\frac{w}{stx},q^{1-\frac{k+j}{2}}\frac{wx}{st};q)}
{w^{2k}x^{2j}(q^{\frac{k+j}{2}}stwx, q^{\frac{k-j}{2}}\frac{stw}{x},
q^{\frac{-k+j}{2}}\frac{stx}{w},q^{-\frac{k+j}{2}}\frac{st}{wx};q) }
\\ \nonumber && \makebox[2em]{} \times
\frac{ (q^{\frac{1+l+j}{2}}\frac{s}{yx}, q^{\frac{1+l-j}{2}}\frac{sx}{y},
q^{\frac{1-l+j}{2}}\frac{sy}{x},q^{\frac{1-l-j}{2}}syx;q)}
{y^{2l}x^{2j}(q^{\frac{1+l+j}{2}}\frac{yx}{s}, q^{\frac{1+l-j}{2}}\frac{y}{sx},
q^{\frac{1-l+j}{2}}\frac{x}{sy},q^{\frac{1-l-j}{2}}\frac{1}{syx};q)}
\\  && \makebox[4em]{}
= D(t;y,l;w,k)M(st)_{w,k;x,j} D(s;y,l;x,j),
\end{eqnarray}
which proves the required identity.

We note that the derived solution of the star-triangle relation
resembles structurally a different solution obtained in \cite{kels}.
We stress that the parameters $y$ and $l$ are dummy variables in this construction, i.e.
at each step of the walk along the lattice of Bailey pairs one can introduce
further new parameters $y,l\to y',l'\to\ldots$.

\section{Coxeter relations and the vertex type Yang-Baxter equation}

Consider elementary transposition operators $s_j,\, j=1,\ldots,5,$ acting on six
parameters $\mathbf{t}=(t_1,\ldots,t_6)$:
\begin{equation}
s_j(\ldots, t_j,t_{j+1},\ldots )=(\ldots,t_{j+1}, t_j,\ldots).
\end{equation}
They generate the permutation group $\mathfrak{S}_6$ characterized by the
Coxeter relations
\begin{equation}
s_j^2=1, \quad s_is_j=s_js_i \
\text{ for } \ |i-j|>1, \quad
s_js_{j+1}s_j=s_{j+1}s_js_{j+1}.
\end{equation}

Define now five operators $\mathrm{S}_j(\mathbf{t}), \, j=1,\ldots, 5,$ acting on
the three-index functions of three
complex variables $f_{n_1,n_2,n_3}(z_1,z_2,z_3)$:
\begin{eqnarray*} && 
[\mathrm{S}_1(\mathbf{t})f]_{n_1,n_2,n_3}(z_1,z_2,z_3):=
M(t_1/t_2)_{z_1,n_1;z,m}f_{m,n_2,n_3}(z,z_2,z_3), \quad
\\ && 
[\mathrm{S}_2(\mathbf{t})f]_{n_1,n_2,n_3}(z_1,z_2,z_3):=
D(t_2/t_3;z_1,n_1;z_2,n_2)f_{n_1,n_2,n_3}(z_1,z_2,z_3).
\\ && 
[\mathrm{S}_3(\mathbf{t})f]_{n_1,n_2,n_3}(z_1,z_2,z_3):=
M(t_3/t_4)_{z_2,n_2;z,m}f_{n_1,m,n_3}(z_1,z,z_3),
\\ && 
[\mathrm{S}_4(\mathbf{t})f]_{n_1,n_2,n_3}(z_1,z_2,z_3):=
D(t_4/t_5;z_2,n_2;z_3,n_3)f_{n_1,n_2,n_3}(z_1,z_2,z_3),
\\ && 
[\mathrm{S}_5(\mathbf{t})f]_{n_1,n_2,n_3}(z_1,z_2,z_3):=
M(t_5/t_6)_{z_3,n_3;z,m}f_{n_1,n_2,m}(z_1,z_2,z),
\end{eqnarray*}
We stress that all these operators depend on the ratios of parameters,
$\mathrm{S}_j(\mathbf{t})=\mathrm{S}_j(t_j/t_{j+1})$.
Let us prove that for an appropriate space of test functions
the operators $\mathrm{S}_j$ generate the group $\mathfrak{S}_6$,
provided their sequential action is defined via a cocycle condition
$\mathrm{S}_j\mathrm{S}_k:=\mathrm{S}_j(s_k(\mathbf{t}))\mathrm{S}_k(\mathbf{t})$.
For this it is necessary to verify the Coxeter relations
\begin{equation}
\mathrm{S}_j^2=1, \quad \mathrm{S}_i\mathrm{S}_j=\mathrm{S}_j\mathrm{S}_i \
\text{ for } \ |i-j|>1, \quad
\mathrm{S}_j\mathrm{S}_{j+1}\mathrm{S}_j
=\mathrm{S}_{j+1}\mathrm{S}_j\mathrm{S}_{j+1}.
\label{coxeter}\end{equation}

Indeed, the latter relations
are equivalent to algebraic properties of the Bailey lemma entries,
in complete analogy with the elliptic hypergeometric case \cite{DS}.
It is sufficient to establish them for $\mathrm{S}_1$ and $\mathrm{S}_2$, others will
follow by the symmetry. So, we have
\begin{equation}
\mathrm{S}_2^2=\mathrm{S}_2(s_2\mathbf{t})\mathrm{S}_2(\mathbf{t})
=D(t_3/t_2;z_1,n_1;z_2,n_2)D(t_2/t_3;z_1,n_1;z_2,n_2)=1.
\end{equation}
A substantially more complicated relation is needed for $\mathrm{S}_1$:
\begin{eqnarray}  \label{chir1} &&  \makebox[-2em]{}
[\mathrm{S}_1^2f]_{n}(x)=[\mathrm{S}_1(s_1\mathbf{t})\mathrm{S}_1(\mathbf{t})f]_{n}(x)
=M(t^{-1})_{x,n;z,m} M(t)_{z,m;y,j}f_{j}(y)
\\ \nonumber &&
= \sum_{j\in\mathbb{Z}}\int [d_jy] \; f_{j}(y) (1-t^2)(1-t^{-2})\;
\\ \nonumber && \qquad \times
\sum_{m\in\mathbb{Z}}\int [d_mz] \; \Gamma_{q}(t^{-1}x^{\pm1},\pm n;z, m)
\Gamma_{q}(ty^{\pm1},\pm j;z, m)
= f_{n}(x),   \quad t=\frac{t_1}{t_2},
\end{eqnarray}
or $\mathrm{S}_1^2=\II$. First, we claim that
$$
M(1)=\II, \quad \text{or} \quad M(1)_{z,m;y,j}f_{j}(y)=f_{m}(z)
$$
for the holomorphic test functions satisfying the reflection
symmetry $f_{-m}(y^{-1})=f_m(y)$. This fact follows from the residue
calculus. For $t\to1$ two pairs of poles approach the integration
contour in $M(t)_{z,m;y,j}f_{j}(y)$ from two sides and pinch it.
To resolve the singularity it is necessary to compute two
residues which leads to the expression $(f_m(z)+f_{-m}(z^{-1}))/2$,
and the reflection symmetry reduces it to one term.
We now substitute in the star-triangle
relation \eqref{STR} the constraint $st=1$. Using the
inversion relation for $D$-function and
$D(1;z_1,n_1;z_2,n_2)=1$, the $D$-terms disappear
on both sides and we obtain $M(t^{-1})M(t)=\II$.

Finally,
\begin{eqnarray} \nonumber &&  \makebox[-2em]{}
\mathrm{S}_1\mathrm{S}_2\mathrm{S}_1=\mathrm{S}_1(s_2s_1\mathbf{t})
\mathrm{S}_2(s_1\mathbf{t})\mathrm{S}_1(\mathbf{t})
=M(\textstyle{\frac{t_2}{t_3}})_{z_1,n_1;z,m}
D(\textstyle{\frac{t_1}{t_3}};z_2,n_2;z,m)  M(\textstyle{\frac{t_1}{t_2}})_{z,m;x,j}
\\  \nonumber &&  \makebox[4em]{}
=\mathrm{S}_2\mathrm{S}_1\mathrm{S}_2=\mathrm{S}_2(s_1s_2\mathbf{t})
\mathrm{S}_1(s_2\mathbf{t})\mathrm{S}_2(\mathbf{t})
\\  &&  \makebox[4em]{}
=D(\textstyle{\frac{t_1}{t_2}};z_1,n_1;z_2,n_2)  M(\textstyle{\frac{t_1}{t_3}})_{z_1,n_1;x,j}
D(\textstyle{\frac{t_2}{t_3}};x,j;z_2,n_2),
\end{eqnarray}
which is precisely the star-triangle relation.

Consider the tensor product of three infinite-dimensional (equal or different) spaces
$\V_1\otimes\V_2\otimes\V_3$  and associate with each space $\V_j$ a pair of variables:
the spectral parameter $u_j$ and the spin variable $g_j$,
respectively. Define R-operators $\mathbb{R}_{ik}(u_i,g_i|u_k,g_k)$
acting in a non-trivial way in the subspace $\V_i\otimes\V_k$ with the unity
operator action in its complement.
The vertex type YBE has the form
\begin{eqnarray}\label{YBvertex} &&
\mathbb{R}_{12} (u_1,g_1|u_2,g_2)\,\mathbb{R}_{13}(u_1,g_1|u_3,g_3)\,
\mathbb{R}_{23}(u_2,g_2|u_3,g_3)
\\ && \makebox[3em]{}
=\mathbb{R}_{23}(u_2,g_2|u_3,g_3)\,\mathbb{R}_{13}(u_1,g_1|u_3,g_3)\,
\mathbb{R}_{12}(u_1,g_1|u_2,g_2).
\nonumber\end{eqnarray}
Actually, the R-operators depend on the difference of spectral parameters,
\begin{equation}
\mathbb{R}_{ik}(u_i,g_i|u_k,g_k)=\mathbb{R}_{ik}(u_i-u_j),
\end{equation}
where we
omitted dependence on the spin variables. Using this notation we can rewrite YBE in the more conventional form
\begin{equation}\label{YBvertex'}
\mathbb{R}_{12} (u-v)\,\mathbb{R}_{13}(u-w)\, \mathbb{R}_{23}(v-w)
=\mathbb{R}_{23}(v-w)\,\mathbb{R}_{13}(u-w)\,\mathbb{R}_{12}(u-v),
\end{equation}
where $u=u_1, v=u_2, w=u_3$.
It is convenient to single out the permutation operators from the R-operator
\begin{equation}
\mathbb{R}_{ik}(u)= \mathbb{P}_{ik}\,\mathrm{R}_{ik}(u),
\end{equation}
where the operator  $\mathbb{P}_{ik}$ interchanges the spaces,
$\mathbb{P}_{ik} (\mathbb{V}_i\otimes\mathbb{V}_k)=\mathbb{V}_k\otimes\mathbb{V}_i$.
Removing these permutation operators from the Yang-Baxter equation \eqref{YBvertex}
yields the relation
\begin{eqnarray} &&  \nonumber
\mathrm{R}_{23}(u_1,g_1|u_2,g_2)\,\mathrm{R}_{12}(u_1,g_1|u_3,g_3)\,
\mathrm{R}_{23}(u_2,g_2|u_3,g_3)\,
\\ && \makebox[3em]{}
=
\mathrm{R}_{12}(u_2,g_2|u_3,g_3)\,\mathrm{R}_{23}(u_1,g_1|u_3,g_3)\,
\mathrm{R}_{12}(u_1,g_1|u_2,g_2),
\label{RRR}\end{eqnarray}
where one sees only two R-operators, $\mathrm{R}_{12}$ and $\mathrm{R}_{23}$.

Let us fix the spaces $\V_j$ as copies of the infinite bilateral sequences of
meromorphic functions $f_j(z),\, j\in\Z$. Then the triple tensor product
of interest takes the form $\V_1\otimes\V_2\otimes\V_3=f_{n_1,n_2,n_3}(z_1,z_2,z_3)$.
Define now the composite operators acting in this space $\mathrm{R}_{12}(\mathbf{t})$,
\begin{eqnarray}\label{R12} &&
\mathrm{R}_{12}(\mathbf{t}) =\mathrm{R}_{12}(t_1,\ldots,t_4)=
\mathrm{S}_2(s_1s_3s_2\mathbf{t})\,
\mathrm{S}_1(s_3s_2\mathbf{t})\,\mathrm{S}_3(s_2\mathbf{t})\,
\mathrm{S}_2(\mathbf{t})
\\ && \makebox[4em]{}
 =\mathrm{S}_2(t_1/t_4)\mathrm{S}_1(t_1/t_3)\mathrm{S}_3(t_2/t_4)\,
\mathrm{S}_2(t_2/t_3),
\nonumber\end{eqnarray}
and $\mathrm{R}_{23}(\mathbf{t})$,
\begin{eqnarray}\label{R23} &&
\mathrm{R}_{23}(\mathbf{t})=\mathrm{R}_{23}(t_3,\ldots,t_6)
=\mathrm{S}_4(s_3s_5s_4\mathbf{t})\,
\mathrm{S}_3(s_5s_4\mathbf{t})\,\mathrm{S}_5(s_4\mathbf{t})\,
\mathrm{S}_4(\mathbf{t})
\\ && \makebox[4em]{}
=\mathrm{S}_4(t_3/t_6)\,
\mathrm{S}_3(t_3/t_5)\,\mathrm{S}_5(t_4/t_6)\,\mathrm{S}_4(t_4/t_5).
\nonumber\end{eqnarray}
Denoting
\begin{equation}
t_{1,2}=e^{-\pi i (u\pm g_1)},\quad t_{3,4}=e^{-\pi i (v\pm g_2)},\quad
 t_{5,6}=e^{-\pi i (w\pm g_3)},
\end{equation}
one can identify
\begin{equation}
\mathrm{R}_{12}(\mathbf{t}) =\mathrm{R}_{12}(u,g_1|v,g_2),
\qquad
\mathrm{R}_{23}(\mathbf{t})=\mathrm{R}_{23}(v,g_2|w,g_3)
\end{equation}
and check that these operators depend only on the difference of spectral parameters
$u-v$ and $v-w$, respectively.

{\bf Theorem}.
The R-operators \eqref{R12} and \eqref{R23} satisfy the vertex type Yang-Baxter
relation \eqref{RRR}.

{\bf Proof}. Substituting the explicit forms of the R-operators into equality
\eqref{RRR}, we come to the relation
\begin{equation}
\mathrm{S}_4\mathrm{S}_3 \mathrm{S}_5 \mathrm{S}_4 \cdot
\mathrm{S}_2\mathrm{S}_1 \mathrm{S}_3\mathrm{S}_2 \cdot
\mathrm{S}_4\mathrm{S}_3 \mathrm{S}_5 \mathrm{S}_4
=
\mathrm{S}_2\mathrm{S}_1 \mathrm{S}_3\mathrm{S}_2 \cdot
\mathrm{S}_4\mathrm{S}_3 \mathrm{S}_5 \mathrm{S}_4 \cdot
\mathrm{S}_2\mathrm{S}_1 \mathrm{S}_3\mathrm{S}_2,
\end{equation}
which is easily checked using only the cubic Coxeter relations
for operators $\mathrm{S}_j$ in complete analogy with the cases
considered in \cite{DM,DS}.

\section{A new two-dimensional solvable lattice model}

Let us apply the operator relation \eqref{STR} to a product of the Kronecker
and Dirac delta-functions which remove integration over the $x$-variable
and summation over the index $j$. This yields the functional star-triangle
relation of the form
\begin{eqnarray}\nonumber
&& \sum_{m\in\Z}\int_{0}^{1} \rho_m(u)\mathrm{W}_{\xi-a}(x,j;u,m)\mathrm{W}_{a+b}(y,j;u,m)
\mathrm{W}_{\xi-b}(w,l;u,m)du
\\ && \makebox[2em]{}
=\chi(a, b)\mathrm{W}_b(x,j;y,k)\mathrm{W}_{\xi-a-b}(x,j;w,l)\mathrm{W}_{a}(y,k;w,l),
\label{astr}\end{eqnarray}
where
\begin{equation}
\mathrm{W}_{a}(x,j;u,m)=\Gamma_q(e^{ 2\pi i(a-\xi\pm x \pm u)}),\quad
 e^{-4\pi\textup{i}\xi}:=q,
\label{weight}\end{equation}
and
\begin{align}
\rho_m(u) & =\frac{(1-q^me^{4\pi iu})(1-q^me^{-4\pi iu})}{2q^m}, \\
\quad \chi(a,b) & = \frac{(qe^{4\pi i a},qe^{4\pi ib}, e^{-4\pi i(a+b)};q)_\infty}
{(e^{-4\pi i a},e^{-4\pi ib}, qe^{4\pi i(a+b)};q)_\infty}.
 \end{align}

We now define a two-dimensional lattice model associated with this relation.
Consider a honeycomb lattice with the spins denoted by labels $x,u,w,$ etc
which seat in vertices. Each spin has a discrete internal degree of freedom
denoted as $m,j,k,l,$ etc (the monopole number). Neighboring spins $(x,j)$
and $(u,m)$ interact along the edges connecting them with the energy
determined by the Boltzmann weight $\mathrm{W}_{a}(x,j;u,m)$.
The function $\rho_m(u)$ describes the self-energy of spins, and
$\xi$ is called the crossing parameter.  In this
picture the ``integration-plus-summation" in the star-triangle relation
\eqref{astr} means computation of the partition function
for an elementary star-shaped cell with contributions coming from all possible
values of the continuous spin $u$ sitting in the central vertex and
all possible values of the magnetic charge $m$. The  honeycomb lattice
can be transformed using the star-triangle relation to triangular
and square lattices.

Compose now $N\times M$ sized two-dimensional square lattice of spins and
associate with each horizontal edge the weight $\mathrm{W}_{a}(x,j;u,m)$
and with the vertical one the weight $\mathrm{W}_{\xi-a}(x,j;u,m)$.
Then the partition function of such homogeneous spin system
with the internal spin energy $\rho_m(u)$ has the form
\begin{equation}
Z=\sum_{\Z^{NM}}\int_{[0,1]^{NM}} \prod_{(ij)}\mathrm{W}_{a}(u_i,m_i;u_j,m_j)
\prod_{(kl)}\mathrm{W}_{\xi-a}(u_k,m_k;u_l,m_l)
\prod_{s}\rho_{m_s}(u_s)du_s,
\end{equation}
where the first product is taken over the horizontal edges $(ij)$,
the second product goes over all vertical edges $(k,l)$,
and the third product (in $s$) is taken over all internal vertices of the
lattice.
Then one can consider the thermodynamical limit of infinite lattice,
$N,M\to \infty$, and look for the free energy per spin $\kappa(a)$
found from the asymptotics
\begin{equation}
Z(a)\stackreb{=}{N,M\to \infty} e^{-NM\kappa(a)}.
\end{equation}
Conjecturally, similar to the models considered in \cite{BMS,BS,astr},
the value of $\kappa(a)$ can be found using the reflection method \cite{bax_refl}.
Namely, one renormalizes the Bolztmann weights
\begin{equation}
\mathrm{\widetilde W}_{a}(x,j;u,m)= \frac{1}{m(a)}\mathrm{W}_{a}(x,j;u,m)
\label{tildeW}\end{equation}
and chooses the multiplier $m(a)$ in such a way that the star-triangle relation
takes the form
\begin{eqnarray}\nonumber
&& \sum_{m\in\Z}\int_{0}^{1} \rho_m(u)\mathrm{\widetilde W}_{\xi-a}(x,j;u,m)
\mathrm{\widetilde W}_{a+b}(y,j;u,m)
\mathrm{\widetilde W}_{\xi-b}(w,l;u,m)du
\\ && \makebox[2em]{}
=\mathrm{\widetilde W}_b(x,j;y,k)\mathrm{\widetilde W}_{\xi-a-b}(x,j;w,l)
\mathrm{\widetilde W}_{a}(y,k;w,l).
\label{astr'}\end{eqnarray}
Then,
\begin{equation}
Z(a)\stackreb{=}{N,M\to \infty} m(a)^{NM}, \quad \text{or}\quad
\kappa(a)=-\log m(a).
\end{equation}
Such a transformation of star-triangle relation requires
\begin{equation}
\frac{m(\xi-a)m(\xi-b)m(a+b)}{m(a)m(b)m(\xi-a-b)}=\chi(a,b),
\end{equation}
which is possible if $m(a)$ satisfies the equation
\begin{equation}
\frac{m(a)}{m(\xi-a)}\frac{(e^{4\pi i(a-\xi)};q)_\infty}{(e^{-4\pi ia};q)_\infty}=1,
\quad \text{or}\quad
m(a+\xi)=\frac{(e^{-4\pi i(a+\xi)};q)_\infty}{(e^{4\pi ia};q)_\infty}m(-a).
\label{ke}\end{equation}

Introduce the following infinite product
\begin{equation}
f(x;p,q)=(x;p,q)_\infty (pqx^{-1};p,q)_\infty,\qquad \frac{f(px;p,q)}{f(x;p,q)}
=\frac{(qx^{-1};q)_\infty}{(x;q)_\infty}.
\end{equation}
We note that this is the product of the numerator and denominator of the
elliptic gamma function.
One has the following inversion relation
\begin{equation}
f(x^{-1};p,q)=f(pqx;p,q).
\end{equation}
Define the composite function
\begin{equation}
\mu (x;p,q)=\frac{f(xp\sqrt{pq};p^2,q)}{f(x\sqrt{pq};p^2,q)}.
\end{equation}
It satisfies the equations
\begin{equation}
\mu (x;p,q)\mu (x^{-1};p,q)=1,\qquad \mu (x;p,q)\mu (p^{-1}x;p,q)=\frac{(x^{-1}p^{1/2}q^{1/2};q)_\infty}
{(xp^{-1/2}q^{1/2};q)_\infty}.
\end{equation}
Using these relations we can set
\begin{equation}
m(a)=\mu (e^{4\pi i a};q,q)=\frac{(q^2e^{4\pi i a},qe^{-4\pi i a};q,q^2)_\infty}
{(qe^{4\pi i a},q^2e^{-4\pi i a};q,q^2)_\infty}
\end{equation}
and see that this function satisfies the unitarity condition
\begin{equation}
m(-a)=\frac{1}{m(a)}
\end{equation}
and the key starting equation \eqref{ke}.
So, $-\log m(a)$ provides the explicit expression for the free energy per spin
of the discussed two-dimensional ``spin" model. For the model with the
Boltzmann weights \eqref{tildeW} the free energy is equal to zero.

\section{Star-star relations and an IRF model Boltzmann weight}

We consider the simplest consequence of the Bailey chain of identities
for sums of $q$-hypergeometric integrals described above
following the elliptic hypergeometric pattern \cite{spi:bailey}.
For this we use the evident explicit Bailey pair, following from the
integration formula \eqref{qbeta}. Namely, let us choose
\begin{equation}
\alpha_m(z,t)=\prod_{j=1}^4\Gamma_q(a_j,n_j;z,m),
\end{equation}
where $a_j$ are arbitrary parameters.
Substituting this expression into the integral transformation \eqref{BP},
imposing the constraint  $\sum_{j=1}^4n_j=0$,  and choosing
$t^2=q\prod_{j=1}^4a_j^{-1}$,
we derive from the Rosengren identity that
\begin{eqnarray}\nonumber &&
\beta_n(x;t)=
\frac{1}{x^{4n}\prod_{j=1}^4a_j^{2n_j}}
\prod_{1\leq j<k\leq 4} \frac{(q^{1+\frac{n_j+n_k}{2} }a_j^{-1}a_k^{-1};q)_\infty}
{(q^{ \frac{n_j+n_k}{2} }a_ja_k;q)_\infty}
\\ &&  \makebox[4em]{} \times
\prod_{j=1}^4\frac{(q^{1+\frac{n_j+n}{2}}a_j^{-1}t^{-1}x^{-1},
q^{1+\frac{n_j-n}{2}}a_j^{-1}t^{-1}x;q)_\infty}
{(q^{\frac{n_j+n}{2}}a_jtx,q^{\frac{n_j-n}{2}}a_jtx^{-1};q)_\infty}.
\end{eqnarray}

We now take definitions of the Bailey lemma entries \eqref{aBP} and  \eqref{bBP} and
substitute them into the relation $\beta_k'(w;st)=M(st)_{w,k;x,j} \alpha_j'(x;st)$.
This yields the following explicit symmetry transformation law
\begin{eqnarray}
V(\underline{a},\underline{n};q)=\frac{V(\underline{\tilde a},\underline{n};q)}{\prod_{j=1}^8a_j^{n_j}}
\prod_{1\leq j<k\leq 4} \frac{(q^{1+\frac{n_j+n_k}{2} }a_j^{-1}a_k^{-1},
q^{1+\frac{n_{j+4}+n_{k+4}}{2} }a_{j+4}^{-1}a_{k+4}^{-1};q)_\infty}
{(q^{ \frac{n_j+n_k}{2} }a_ja_k,q^{ \frac{n_{j+4}+n_{k+4}}{2} }a_{j+4}a_{k+4};q)_\infty},
\label{E7trafo}\end{eqnarray}
where
\begin{equation}
V(\underline{a},\underline{n};q):=\sum_{m\in\mathbb{Z}}\int_{\mathbb T}\prod_{j=1}^8
\Gamma_q(a_j,n_j;z,m)[d_mz],\qquad \prod_{j=1}^8a_j=q^2,\quad \sum_{j=1}^8n_j=0
\label{Vfunction}\end{equation}
and the following notation for the parameters is used
\begin{equation}
a_{5,6}=stw^{\pm1}, \quad n_{5,6}=\pm k,\quad
a_{7,8}=q^{1/2} s^{-1}y^{\pm1}, \quad n_{7,8}=\pm l
\end{equation}
as well as
\begin{equation}
\tilde a_j=ta_j,\quad j=1,2,3,4,\qquad \tilde a_j=t^{-1}a_j,\quad j=5,6,7,8.
\end{equation}
Remind also the constraint $t^2\prod_{j=1}^4a_j=q$.

{\bf Conjecture}.
Let us take the $V$-function, whose parameters $a_j, n_j$
satisfy only the balancing conditions indicated in the definition \eqref{Vfunction}
and an additional constraint $\sum_{j=1}^4 n_j=0$.
Then we conjecture that it satisfies the symmetry transformation
\eqref{E7trafo}, where
\begin{equation}
\left\{
\begin{array}{cl}
\tilde a_j =\varepsilon a_j,&   j=1,2,3,4  \\
\tilde a_j = \varepsilon^{-1} t_j, &    j=5,6,7,8
\end{array}
\right.;
\quad \varepsilon=\sqrt{\frac{q}{a_1a_2a_3a_4}}=\sqrt{\frac{a_5a_6a_7a_8}{q}}.
\label{partrafo}\end{equation}

Indeed, using the relation
\begin{equation}
\frac{ (q^{1-m/2}z^{-1} ;q)_\infty} {( q^{-m/2} z;q)_\infty }
=\frac{q^{m/2}}{(-z)^m} \frac{ (q^{1+m/2}z^{-1} ;q)_\infty} {( q^{+m/2} z;q)_\infty }, \quad m\in\Z,
\label{identity}\end{equation}
one can verify that a repetition of the transformation \eqref{E7trafo}, \eqref{partrafo}
returns back the original $V$-function, i.e. we deal with a reflection.
The map $a_j\to \tilde a_j$ is the key reflection extending the
Weyl group $S_8$ of the root system $A_7$ to the Weyl
group of the exceptional root system $E_7$. However, because of the
presence of integers $n_j$ and the constraint $\sum_{j=1}^4n_j=0$
we do not have the full $W(E_7)$ symmetry of the $V$-function yet.
Interestingly, even in this reduced case the Bailey chains
techniques yields the symmetry transformation \eqref{E7trafo} only
when a pair of integers is forced to take particular
values $n_i+n_j=n_k+n_l=0$, $i\neq j\neq k\neq l$,
which contrasts with the elliptic hypergeometric $V$-function case \cite{S2,S3}.

Consider a $2d$ checkerboard lattice \cite{bax_ss}
where each ``black" site has four
``white" neighbours and, vice versa, each ``white" site has four
``black" neighbours. Ascribe to each edge connecting the white and black
sites the Boltzmann weight $\mathrm{W}_{\alpha_i}$ \eqref{weight}
with arbitrary parameters $\alpha_i$ subject to the constraint $\sum_{j=1}^4\alpha_j=2\xi$.
An IRF model is obtained when we integrate out the one-color lattice spins.
The Boltzmann weight of the corresponding elementary ``cell" containing  four
vertices determines the energy of this square face. It is given obviously
by a special case of the general $V$-function introduced above
when all integer variables $n_j$ are paired by the relation $n_{2i-1}+n_{2i}=0$.
Then, completely similarly to \cite{astr}, the symmetry transformation \eqref{E7trafo}
has now the interpretation as a star-star relation \cite{bax_ss}.
As shown by Baxter \cite{bax_ss2} knowledge of the star-star relations
automatically leads to the YBE for IRF models.

\section{IRF Yang-Baxter equation with spectral parameter}

The Yang-Baxter equation for IRF models (or SOS-type YBE)
\cite{SOS1,SOS2} associated with $3d$ superconformal indices has the following form
\begin{eqnarray} \nonumber &&
\sum_{H \in \mathbb Z} \int [d_H h] \; R_{t_{41}t_{63}}\left( \begin{array}{cc}
  {a, A} & {}{b, B}\\ {}{h, H} &
  {}{c, C}\end{array}\right) \; R_{t_{63}t_{25}}\left(\begin{array}{cc}
    {}{c, C} & {}{d, D}\\ {}{h, H} &
    {}{e, E}\end{array}\right)
\\ \nonumber  && \makebox[2em]{} \times
 R_{t_{25}t_{41}}\left(\begin{array}{cc}
      {}{e, E} & {}{f, F}\\ {}{h, H} &
      {}{a, A}\end{array}\right)
= \sum_{H \in \mathbb Z} \int [d_H h] \; R_{t_{63}t_{25}}\left(\begin{array}{cc}
        {}{b, B} & {}{h, H}\\ {}{a, A} &
        {}{f, F}\end{array}\right)
\\ \label{SOSYBE}    &&  \makebox[4em]{} \times
 R_{t_{25}t_{41}}\left(\begin{array}{cc}
          {}{d, D} & {}{h, H}\\ {}{c, C} &
          {}{b, B}\end{array}\right) \; R_{t_{41}t_{25}}\left(\begin{array}{cc}
            {}{f, F} & {}{h, H}\\ {}{e, E} &
            {}{d, D}\end{array}\right),
\end{eqnarray}
where we introduced for convenience the shorthand notation for spectral
parameters $t_{ij}=(t_i,t_j)$. The following statistical weight satisfies this equation
\begin{align} \nonumber
R_{(m,l)(n,r)}\left(\begin{array}{cc}
  {}{a, A} & {}{b, B}\\ {}{d, D} &
  {}{c, C}\end{array}\right) & \ =
\frac{(q^{\frac23} (n/l)^{-2}, q^{\frac23} (r/m)^{-2} ; q)_\infty}{(q^{\frac13}
  (n/l)^2,q^{\frac13} (r/m)^2;q)_\infty}
\sum_{k  \in \mathbb Z}  \int [d_k  z]  \\ \nonumber
 & \times \Gamma_q(q^{\frac13} \frac{l}{n} a^{\pm 1},\pm A ; z,k )  \Gamma_q(q^{\frac16} \frac{r}{l} b^{\pm 1},\pm B  ;z,k )  \\ \label{RSOS}
 & \times    \Gamma_q(q^{\frac13} \frac{m}{r} c^{\pm 1},\pm C  ; z,k )
\Gamma_q( q^{\frac16} \frac{n}{m} d^{\pm 1} ,\pm D ;z,k ).
\end{align}
It is substantially equal to the $V$-function \eqref{Vfunction}
with particular constraints on the integers $\underline{n}=(\pm A, \pm B, \pm C, \pm D)$.

For showing that function (\ref{RSOS}) describes a solution of equation
\eqref{SOSYBE} we use a special case of identity \eqref{qbeta} associated
with the star-triangle relation
\begin{eqnarray} \nonumber
&& 
 \sum_{m \in \mathbb Z} \int [d_m z] \Gamma_q(q^{\frac16} t/s a^{\pm 1},\pm A ;z,m)
 \Gamma_q(q^{\frac16} s/r b^{\pm 1},\pm B  ;z,m) \Gamma_q(q^{\frac16} r/t c^{\pm 1},\pm C  ;z,m)
\\   \nonumber  &&
 \ = \  \frac{(q^\frac23 (t/s)^{-2}, q^\frac23 (s/r)^{-2},
q^\frac23 (r/t)^{-2};q)_\infty}{(q^\frac13 (t/s)^2,
q^\frac13 (s/r)^2, q^\frac13 (r/t)^2;q)_\infty}
\Gamma_q(q^{\frac13} t/r a^{\pm 1},\pm A ; b,B)  \\
&& \qquad \times \;
 \Gamma_q(q^{\frac13} r/s  c^{\pm 1},\pm C ; a,A)
\Gamma_q(q^{\frac13} s/t b^{\pm 1},\pm B  ; c,C).
\label{SOSSTR}\end{eqnarray}
We now form the following composite function defined by 6 integrations and
6 discrete summations
\begin{eqnarray} \nonumber
&& \sum_{m_i \in \mathbb Z} \int \prod_{i=1}^6 [d_{m_i}z] \;
\Gamma_q(q^{\frac16} t_1/t_5 f^{\pm 1},\pm F ;z_6,m_6) \Gamma_q(q^{\frac16} t_6/t_1 z_6^{\pm 1},\pm m_6 ;z_1,m_1) \\ \nonumber
&& \qquad \qquad  \times \; \Gamma_q(^{\frac16} t_2/t_6 a^{\pm 1},\pm A ;z_1,m_1) \;
 \Gamma_q(q^{\frac16} t_1/t_2 z_2^{\pm 1},\pm m_2 ;z_1,m_1) \\ \nonumber
&& \qquad \qquad  \times \; \Gamma_q(q^{\frac16} t_3/t_1 b^{\pm 1},\pm B  ;z_2,m_2) \Gamma_q(q^{\frac16} t_2/t_3 z_3^{\pm 1},\pm m_3 ;z_2,m_2) \;\\ \nonumber
&& \qquad \qquad  \times \; \Gamma_q(q^{\frac16} t_4/t_2 c^{\pm 1},\pm C  ;z_3,m_3) \Gamma_q(q^{\frac16} t_3/t_4 z_4^{\pm 1},\pm m_4 ;z_3,m_3) \\ \nonumber
&& \qquad \qquad \times \; \Gamma_q(q^{\frac16} t_5/t_3 d^{\pm 1},\pm D ;z_4,m_4)  \Gamma_q(q^{\frac16} t_4/t_5 z_5^{\pm 1},\pm m_5 ;z_4,m_4) \; \\
 && \qquad \qquad \times \; \Gamma_q(q^{\frac16}t_6/t_4 e^{\pm 1},\pm E ;z_5,m_5)
\Gamma_q(q^{\frac16} t_5/t_6 z_6^{\pm 1},\pm m_6 ;z_5,m_5).
\end{eqnarray}
Then we integrate over $z_1$, $z_3$, and $z_5$ and sum over $m_1$, $m_3$,
and $m_5$, i.e. use the star-triangle relation \eqref{SOSSTR} for the
expressions indicated in the square brackets below
\begin{eqnarray} \nonumber
&& \sum_{m_2,m_4,m_6 \in \mathbb Z} \int  [d_{m_2} z] [d_{m_4} z] [d_{m_6}z] \; \Gamma_q(q^{\frac16} t_1/t_5 f^{\pm 1},\pm F ;z_6,m_6) \\ \nonumber
&& \qquad \qquad  \qquad \times \; \Gamma_q(q^{\frac16} t_3/t_1 b^{\pm 1},\pm B  ;z_2,m_2) \Gamma_q(q^{\frac16} t_5/t_3 d^{\pm 1},\pm D ;z_4,m_4) \\ \nonumber
&& \qquad \qquad  \qquad  \times \Big[ \sum_{m_1 \in Z} \int  [d_{m_1} z] \; \Gamma_q(q^{\frac16} t_6/t_1 z_6^{\pm 1},\pm m_6 ;z_1,m_1) \\ \nonumber
&&\qquad \qquad  \qquad \times \; \Gamma_q(q^{\frac16} t_2/t_6 a^{\pm 1},\pm A ;z_1,m_1)
 \Gamma_q(q^{\frac16} t_1/t_2 z_2^{\pm 1},\pm m_2 ;z_1,m_1) \Big] \;\\ \nonumber
 && \qquad \qquad  \qquad \times \Big[ \sum_{m_3 \in Z} \int  [d_{m_3}z] \; \Gamma_q(q^{\frac16} t_2/t_3 z_3^{\pm 1},\pm m_3 ;z_2,m_2) \\ \nonumber
 && \qquad \qquad  \qquad \times \; \Gamma_q(q^{\frac16} t_4/t_2 c^{\pm 1},\pm C  ;z_3,m_3) \Gamma_q(q^{\frac16} t_3/t_4 z_4^{\pm 1},\pm m_4 ;z_3,m_3) \Big] \\ \nonumber
&& \qquad \qquad  \qquad  \times \Big[ \sum_{m_5 \in Z} \int  [d_{m_5}z] \; \Gamma_q(q^{\frac16} t_4/t_5 z_5^{\pm 1},\pm m_5 ;z_4,m_4) \; \\ \nonumber
&& \qquad \qquad  \qquad \times \;  \Gamma_q(q^{\frac16} t_6/t_4 e^{\pm 1},\pm E ;z_5,m_5) \Gamma_q(q^{\frac16} t_5/t_6 z_6^{\pm 1},\pm m_6 ;z_5,m_5) \Big].
\end{eqnarray}
As a result, we obtain
\begin{eqnarray} \nonumber && \makebox[-2em]{}
\frac{(q^{\frac23} (t_6/t_1)^{-2}, q^{\frac23} (t_3/t_4)^{-2}, q^{\frac23} (t_1/t_2)^{-2},q^{\frac23} (t_4/t_5)^{-2}, q^{\frac23} (t_2/t_3)^{-2}, q^{\frac23} (t_5/t_6)^{-2}  ;q)_\infty}
{(q^{\frac13} (t_6/t_1)^2, q^{\frac13} (t_3/t_4)^2, q^{\frac13} (t_1/t_2)^2,
q^{\frac13} (t_4/t_5)^2, q^{\frac13} (t_2/t_3)^2, q^{\frac13} (t_5/t_6)^2; q)_\infty}
\\ \nonumber &&  \makebox[-2em]{}
\times \frac{(q^{\frac23} (t_6/t_4)^{-2}, q^{\frac23} (t_4/t_2)^{-2},
q^{\frac23} (t_2/t_6)^{-2}; q)_\infty}
{(q^{\frac13} (t_6/t_4)^2, q^{\frac13} (t_4/t_2)^2, q^{\frac13} (t_2/t_6)^2; q)_\infty}
\sum_{m_2,m_4,m_6 \in \mathbb Z} \int  [d_{m_2}z] [d_{m_4}z] [d_{m_6}z] \;
 \\ \nonumber &&  \makebox[-2em]{}
\times \;
\Gamma_q(q^{\frac16} \frac{t_1}{t_5} f^{\pm 1},\pm F ;z_6,m_6) \Gamma_q(q^{\frac13} \frac{t_6}{t_5} e^{\pm 1},\pm E ; z_4,m_4)
\Gamma_q(q^{\frac13} \frac{t_5}{t_4} e^{\pm 1},\pm E ; z_6,m_6)
\\ \nonumber &&  \makebox[-2em]{}
\times \; \Gamma_q(q^{\frac13} \frac{t_2}{t_1} a^{\pm 1},\pm A ; z_6,m_6)
  \Gamma_q(q^{\frac13} \frac{t_1}{t_6} a^{\pm 1},\pm A ; z_2,m_2)
\Gamma_q(q^{\frac16} \frac{t_3}{t_1} b^{\pm 1},\pm B  ;z_2,m_2)
\\ \nonumber  &&  \makebox[-2em]{}
\times \;   \Gamma_q(q^{\frac13} \frac{t_4}{t_3} c^{\pm 1},\pm C  ; z_2,m_2)
\Gamma_q(q^{\frac13} \frac{t_3}{t_2} c^{\pm 1},\pm C  ; z_4,m_4)
\Gamma_q(q^{\frac16} \frac{t_5}{t_3} d^{\pm 1},\pm D ;z_4,m_4)
 \\ &&  \makebox[-2em]{}
 \times \Big[ \Gamma_q(q^{\frac13} \frac{t_6}{t_2} z_6^{\pm 1},\pm m_6 ;z_2,m_2)
\Gamma_q(q^{\frac13} \frac{t_2}{t_4} z_4^{\pm 1},\pm m_4 ; z_2,m_2)
\Gamma_q(q^{\frac13} \frac{t_4}{t_6} z_6^{\pm 1},\pm m_6 ; z_4,m_4) \Big].
 \nonumber\end{eqnarray}
Finally, we apply the inverse triangle-star relation to the last line product of
$\Gamma_q$-functions in the square brackets and obtain the left-hand side expression
in equation \eqref{SOSYBE}.
The right-hand side expression of this IRF YBE is obtained after performing
first the integrations over $z_2,z_4,z_6$ and summations over $m_2, m_4, m_6$
and an application of a similar triangle-star transformation.

\section{The $3d$ superconformal index and duality}

In this section we briefly review some necessary details about superconformal
index of three--dimensional supersymmetric theories with four supercharges
(${\mathcal N}=2$ theories). Here we mainly follow the references
\cite{Imamura:2011su,Kapustin:2011jm,Krattenthaler:2011da}.

The superconformal index first was proposed for four-dimensional theories
\cite{Romelsberger1,Kinney} and later extended to other
dimensions. Three--dimensional index was computed  using localization
technique by Kim \cite{Kim:2009wb} for ABJM theory and it was generalized
to ${\mathcal N}=2$ theories by Imamura and Yokoyama \cite{Imamura:2011su}
(with a topological symmetry contribution amendment pointed out
in \cite{Krattenthaler:2011da}).
The superconformal index of three--dimensional ${\mathcal N}=2$
superconformal field theory is a twisted partition
function defined on $S^2 \times S^1$
\cite{Bhattacharya:2008bja,Kim:2009wb,Imamura:2011su}:
\begin{equation}
I(x,t)=\text{Tr} \left[ (-1)^\text{F} \exp (-\beta\{Q, Q^\dagger \})
x^{\Delta+j_3}\prod_j t_j^{F_j} \right],
\end{equation}
where F is the fermion number, $\Delta$ is the energy, $j_3$ is the third component
of the angular momentum around $S^2$, and $F_j$ are the Cartan
generators of the global flavor symmetry.  The trace is taken over the
Hilbert space of the theory. Here, $Q$ is a supersymmetric charge with
quantum numbers $\Delta=\frac12$ and $j_3=-\frac12$ and the $R$-charge
is normalized in a such way that $Q$ has $R$-charge equal to $1$.
The supercharges $Q^\dagger = S$ and $Q$
satisfy the following anti-commutation relation (the full algebra
can be found in many papers, for instance, in \cite{Dolan:2008vc})
\begin{equation}
 2 {\mathcal H} = \{Q, S\}=\Delta-R-j_3,
\end{equation}
where $R$ is the R-charge.
Only the BPS states satisfying the bound ${\mathcal H} =0 $ contribute to the index,
therefore the index is $\beta$-independent.

Using the localization technique \cite{Pestun:2007rz} the superconformal index
can be computed exactly \cite{Kim:2009wb,Imamura:2011su}, and it reduces
to the following matrix integral
\begin{align} \nonumber
I(x,t)& = \sum_{m\in \mathbb{Z}} \int \frac{1}{|W_m|}
e^{-S^{(0)}_{CS}}e^{ib_0} x^{\epsilon_0} \prod_j^{\text{rank} F} t_{j}^{q_{0j}} \\
& \qquad \qquad \times
\exp\left[\sum^\infty_{n=1}\frac{1}{n} \text{ind}(z^{n},
t^n,x^n; m)\right] \; d\mu_G(z) \;.
\label{3dsci}\end{align}
Let us unpack this expression. The summation is over magnetic
fluxes on two-sphere which appears in the localization procedure as a
contribution of monopoles.  The $d\mu_G(z)$ is the Haar measure of the gauge group $G$. The prefactor $|W_m|= \prod_{i=1}^k ( \text{rank}G_i)!$
is the order of the Weyl group of $G$ which is ``broken'' by the monopoles
to the product $G_1 \times G_2 \times \dots  \times G_k$. If the theory has the
Chern-Simons term it contributes to the index as
\begin{equation}
S^{(0)}_{CS} \ = \ \frac{ik}{4 \pi} \int tr_{CS} (A^{(0)} dA^{(0)}
-\frac{2 i} {3} A^{(0)} A^{(0)} A^{(0)}) \ = \ i \; \text{tr}_{CS} (g m) ,
\end{equation}
where $\text{tr}_{CS}$ stands for the trace including the Chern-Simons levels, $g$ runs over the maximal torus of the gauge group and $m$ takes values in the Cartan of the gauge group and parametrizes magnetic monopole charges. There is also the one-loop correction to the Chern-Simons term
\begin{equation}
b_0 \ = \ -\frac{1}{2} \; \sum_\Phi\sum_{\rho\in R_\Phi}|\rho(m)|\rho(g) \;,
\end{equation}
where $\sum_\Phi$ and $\sum_{\rho\in R_\Phi}$ represent summations over all
chiral multiplets and all weights of the representation $R_\Phi$ of the gauge group. The term
$q_{0j}$ is the zero-point contribution to the energy,
\begin{equation}
q_{0j}(m) = -\frac{1}{2} \sum_\Phi \sum_{\rho\in R_\Phi} |\rho(m)| f_j (\Phi),
\end{equation}
and $\epsilon_0$ is the Casimir energy of the vacuum state on two-sphere with
magnetic flux $m$,
\begin{equation}
\epsilon_0(m) = \frac{1}{2} \sum_\Phi (1-r_\Phi) \sum_{\rho\in
R_\Phi} |\rho(m)|
- \frac{1}{2} \sum_{\alpha \in G} |\alpha(m)| \;,
\end{equation}
where $\sum_{\alpha\in G}$ is the sum over all roots of $G$ and $r_\Phi$ is
the $R$-charge of the chiral multiplet. The single letter index
$ \text{ind}(z,t,x;m)$ gets contributions from chiral and vector multiplets
\begin{eqnarray} &&
\text{ind}(z=e^{ig_j},t,x;m)  = -\sum_{\alpha\in G} e^{i\alpha(g)} x^{|\alpha(m)|}
\\ \nonumber  && \makebox[-2em]{}
 + \sum_\Phi \sum_{\rho\in R_\Phi} \left[
e^{i\rho(g)}  t_{j}^{f_{j}}
\frac{x^{|\rho(m)|+r_\Phi}}{1-x^2}  -  e^{-i\rho(g)} 
t_{j}^{-f_{j}} \frac{x^{|\rho(m)|+2-r_\Phi}}{1-x^2}
\right].
\nonumber \end{eqnarray}
The single particle index enters the full superconformal index \eqref{3dsci}
via the ``plethystic exponential"
\cite{Benvenuti:2006qr,Feng:2007ur}
\begin{equation}
\exp \bigg( \sum_{n=1}^\infty \frac{1}{n} \text{ind} ( z^n, t^n, x^n; m) \bigg) \;.
\end{equation}

The three-dimensional superconformal index can be written in terms of sums of basic
hypergeometric integrals, see e.g.
\cite{Krattenthaler:2011da,Kapustin:2011jm,Gahramanov:2013rda,Gahramanov:2014ona}.
For instance, let us consider the ${\mathcal N}=2$ theory with $U(N)$ gauge group.
Then the chiral multiplet $\Phi$ with $R$-charge $r_\Phi$ in the fundamental representation of the gauge group contributes to the index as
\begin{equation} \label{chiral}
\prod_{j=1}^{\text{rank}\, F}\prod_{i=1}^{\text{rank}\, G}
\frac{(x^{2-r_\Phi+|m_i|} t_j^{-1}z_i^{-1}; x^2)_{\infty}}
{(x^{r_\Phi+|m_i|} t_jz_i; x^2)_{\infty}} \;,
\end{equation}
and the corresponding vector superfield contributes as
\begin{equation}
x^{-\sum_{1\leq i<j\leq N} |m_i-m_j|} \prod_{i,j=1,\ldots, N,\, i\neq j}
(1-\frac{z_i}{z_j} x^{|m_i-m_j|}) \;.
\end{equation}

Our main object of interest is the so-called generalized superconformal index which includes
integer parameters corresponding to global symmetries. In \cite{Kapustin:2011jm}
Kapustin and Willett pointed out that it is possible to generalize the superconformal
index of $3d$ ${\mathcal N}=2$ theory by considering a non--trivial
background gauge field coupled to the global symmetries of the theory. Then the
superconformal index includes new discrete parameters for global
symmetries (one can obtain this expression using the localization
technique \cite{Drukker:2012sr}). For instance, the contribution of the chiral
multiplet (\ref{chiral}) in this case gets the following form
\begin{equation}
\prod_{j=1}^{\text{rank}\, F}\prod_{i=1}^{\text{rank}\, G} \frac{(x^{2-r_\Phi+|m_i|+  n_j}
t_j^{-1}z_i^{-1}; x^2)_{\infty}}{(x^{r_\Phi+|m_i|+n_j} t_jz_i; x^2)_{\infty}} \;,
\end{equation}
where the parameters $n_j$ are new discrete variables, and the contribution
of gauge fields remains the same. The
general expression for such an index has the following form
\begin{align} \nonumber
I(\underline{t},\underline{n}; x) = \sum_{m_k\in\mathbb{Z}}\frac{1}{|W_m|}
\int \prod_{k=1}^{\text{rank}\, G} \frac{dz_k}{2 \pi i z_k} Z_{gauge} (z_k,m_k; x^2) \\
\qquad \qquad \times \prod_{\Phi} Z_{\Phi}(z_k,m_k; t_a, n_a; x^2).
\end{align}
We do not write the contribution of the Chern--Simons term, since
we consider theories without this term.

We now want to describe the two-dimensional solvable lattice models discussed
above in the context of supersymmetric dualities for $3d$ ${\mathcal N}=2$ supersymmetric gauge
theories. The duality we study is very similar to the initial Seiberg duality
for ${\mathcal N}=1$ four-dimensional supersymmetric quantum chromodynamics.
The following two theories are dual to each other \cite{Ros2}:
\begin{itemize}
    \item \textbf{Theory A}: $SU(2)$ gauge group with $N_f=6$ flavors, chiral
multiplets in the fundamental representation of the flavor group $SU(6)$ and in the
fundamental representation of the gauge group.\\[-0.2cm]

    \item \textbf{Theory B}: without gauge degrees of freedom and the chiral fields
(gauge-invariant ``mesons'') in the 15-dimensional totally antisymmetric tensor
representation of the flavor group.
\end{itemize}
More precisely, the first interacting gauge fields theory flows in the
infrared limit to the second one.
This duality was considered in \cite{Teschner:2012em}. The authors
calculated the three--dimensional ellipsoid partition functions
for dual theories by applying the reduction procedure of \cite{DSV}
to the models considered in \cite{SV1}.

\begin{figure}[htbp]
    \centering
        \includegraphics[trim=4.3cm 20cm 0.5cm 4cm, width=1.30\textwidth]{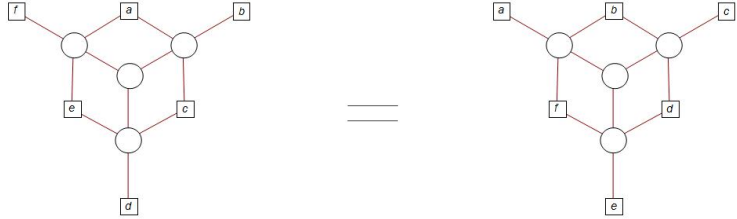}
    \caption{Duality of quiver diagrams.}
    \label{fig:somthing}
\end{figure}

The ordinary superconformal index of the ``theory A'' with
enhanced symmetry  was presented in \cite{DG} (see also \cite{Gahramanov:2013xsa} for
the $N_f=4$ case and \cite{Gahramanov:2013rda,Gahramanov:2014ona} for the similar
theory with the broken gauge group). The duality between theories A and B
leads to the equality of corresponding superconformal indices
expressed by the following $q$-hypergeometric identity \cite{Ros2} (after denoting $x^2=q$)
\begin{align} \nonumber
& \sum_{m\in\mathbb{Z}}\int_{\mathbb T} \; q^{-|m|} \; \prod_{j=1}^6
\frac{(q^{1+\frac{n_j}{2}+\frac{|m|}{2} }\frac{1}{a_jz},q^{1+\frac{n_j}{2}+\frac{|m|}{2} }\frac{z}{a_j};q)_\infty}
{(q^{\frac{n_j}{2}+\frac{|m|}{2} }a_jz,q^{\frac{n_j}{2}+\frac{|m|}{2} }\frac{a_j}{z};q)_\infty}
(1-q^{|m|} z^2)(1-q^{|m|} z^{-2}) \frac{dz}{2\pi iz} \\ \label{indfordual}
& \qquad \qquad \qquad =\frac{1}{\prod_{j=1}^6 a_j^{n_j}}
\prod_{1\leq j<k\leq 6} \frac{(q^{1+\frac{n_j}{2} +\frac{n_k}{2}}a_j^{-1}a_k^{-1};q)_\infty}
{(q^{\frac{n_j}{2}+\frac{n_k}{2}}a_ja_k;q)_\infty},
\end{align}
with the balancing condition
\begin{equation}
\prod_{j=1}^6 a_j=q, \;\; \text{and} \;\; \sum_{j=1}^{6}n_j=0 \;.
\end{equation}
This condition is imposed by the effective superpotential $W=\eta X$
for the theory A, where $X$ is a monopole operator and $\eta$ is the
four-dimensional instanton factor, which breaks a part of the symmetry
(for details, see \cite{Aharony:2013dha}).
Using the relation \cite{Dimofte:2011py}
\begin{equation}
 \prod_{i=0}^{\infty} \frac{1-q^{i-\frac12 m+1}z^{-1}}{1-q^{i-\frac12 m} z}
\ = \ (-q^{\frac12})^{\frac12(m+|m|)} z^{-\frac12(m+|m|)} \prod_{i=0}^{\infty}
\frac{1-q^{i+\frac12 |m|+1} z^{-1}} {1-q^{i+\frac12|m|} z}
\end{equation}
one can obtain the $q$-beta sum-integral (\ref{qbetaint}) from (\ref{indfordual}).

Similarly, the full symmetry transformation \eqref{E7trafo} is a consequence
of a duality of two $3d$ theories with $N_f=8$. One can guess that there exist
proper analogs of all elliptic hypergeometric integral identities
described in \cite{S3,SV1,SV3,SV3b} for sums of $q$-hypergeometric integrals
associated with $3d$ dualities. Actually, the latter dualities are easily found
using the reduction of $4d$ superconformal indices to $3d$ partition functions
\cite{DSV} which naturally leads to conjectural equalities
of corresponding $3d$ superconformal indices.

By breaking the flavor symmetry to $SU(2) \times SU(2) \times SU(2)$ in
(\ref{indfordual}) we obtain the star-triangle relation (\ref{SOSSTR}).
Then the expression (\ref{RSOS}) corresponds to the generalized superconformal
index of a $3d$ ${\mathcal N}=2$ theory with the gauge group $G=SU(2)$ and the flavor group
$F=SU(2) \times SU(2) \times SU(2) \times SU(2)$. In this picture, the SOS-type YBE
(\ref{SOSYBE}) is nothing else than the equality of superconformal indices
of two dual $3d$ ${\mathcal N}=2$ supersymmetric quiver gauge theories presented in Fig. 1,
where the boxes correspond to $SU(2)$ flavor subgroups
and the circles represent $SU(2)$ gauge subgroups.

We note that relation \eqref{chir1} describes the chiral symmetry breaking
similarly to the $3d$ partition function case \cite{SV6}. Indeed, it
assumes the following sum-integral evaluation
\begin{align} \nonumber
& \sum_{m\in\mathbb{Z}}\int [d_mz] \; \Gamma_{q}(t^{-1}x^{\pm1},\pm n;z, m)
\Gamma_{q}(ty^{\pm1},\pm j;z, m) \\
& \qquad =\frac{\delta(\phi_y + \phi_x)\delta_{n + j,0}+
\delta(\phi_y - \phi_x)\delta_{n - j,0}}
{q^{-j}(1-q^jy^2)(1-q^jy^{-2})(1-t^2)(1-t^{-2})},
\label{CSBind}\end{align}
where $y=e^{2\pi i \phi_y}$ and $x=e^{2\pi i \phi_x}$ and
$\delta (\phi)$ is the periodic Dirac delta function with period 1,
$\delta(\phi+1)=\delta(\phi)$. On the left-hand side of equality \eqref{CSBind}
we have the  $3d$ superconformal index of a theory with $SU(2)$
gauge group and $N_f=4$ chiral fields with the naive flavor group $SU(2)\times SU(2)$.
However, as follows from the the right-hand side expression, the
true flavor group is $(SU(2)\times SU(2))_{diag}$ and the superconformal
index has, actually, a non-zero support only on the corresponding
subset of fugacities. This is precisely the manifestation of chiral
symmetry breaking in confining theories similar to the $3d$ partition
functions case \cite{SV6}. A more detailed and rigorous consideration
of this relation between indices and spontaneous breaking of global
symmetries is needed, in particular, for the case when one has
originally the full naive $SU(4)$ flavor group which is broken to $SP(4)$ group.

\smallskip

{\bf Acknowledgements}.
The authors are indebted to H. Rosengren for helpful discussions.
This work is supported by the Heisenberg-Landau program and the
Russian foundation for basic research (grant no. 14-01-00474).
The research of IG is supported
in part by the SFB 647 ``Raum-Zeit-Materie. Analytische und Geometrische Strukturen'',
the Research Training Group GK 1504 ``Mass, Spectrum, Symmetry'' and  the
International Max Planck Research School for Geometric Analysis, Gravitation
and String Theory.
After completion of this work we have known that the functional
form of the star-triangle relation of Sect. 5 was considered in \cite{K2}
and a relation of $3d$ superconformal indices to
integrable lattice systems was discussed in \cite{Yagi}.

\end{document}